\documentclass[aps,pra,twocolumn,groupedaddress,showpacs]{revtex4}
\usepackage{epsfig}
\usepackage{graphics}
\usepackage{amssymb}
\usepackage{amsmath}

\usepackage{epsfig}

\begin{document}

\title{Heralding Single Photons Without Spectral Factorability}
\author{Yu-Ping Huang}
\affiliation{Center for Photonic Communication and Computing, EECS Department\\
Northwestern University, 2145 Sheridan Road, Evanston, IL 60208-3118}
\author{Joseph B. Altepeter}
\affiliation{Center for Photonic Communication and Computing, EECS Department\\
Northwestern University, 2145 Sheridan Road, Evanston, IL 60208-3118}
\author{Prem Kumar}
\affiliation{Center for Photonic Communication and Computing, EECS Department\\
Northwestern University, 2145 Sheridan Road, Evanston, IL 60208-3118}

\begin{abstract}
Recent efforts to produce single photons via heralding have relied on creating spectrally factorable two-photon states in order to achieve both high purity and high production rate. Through a careful multimode analysis, we find, however, that spectral factorability is not necessary. Utilizing single-mode detection, a similar or better performance can be achieved with non-factorable states. This conclusion rides on the fact that even when using a broadband filter, a single-mode measurement can still be realized, as long as the coherence time of the triggering photons exceeds the measurement window of the on/off detector.
\end{abstract}

\maketitle
\section{Introduction}
Pure and single photons in well-defined spatiotemporal
modes are a prerequisite for many quantum applications, such as quantum cryptography and linear optical quantum computing
\cite{KniLafMil01,Rev-Qua-cryp}. In order to achieve high collection efficiency, sources of single photons via
heralding have been extensively studied and demonstrated in both waveguides
\cite{Heralded-Single-Photon-SPDC86,Heralding-SPDC86,Single-Photon-PDC-99,Single-Photon-PDC01,Single-Photon-PDC07,Herald-Single-Photon-08}
and fibers
\cite{Single-Photon-PCF05,Single-Photon-PCF08,Single-Photon-Fiber09,Single-Photon-fiber09-2}.
To herald a pure photon, it is critical that the detection of the
triggering photon of a spontaneously generated pair (called the ``signal'') does not impose any
distinguishing information on the heralded photon (the ``idler'').
Such distinguishability usually arises from the spectral correlation between the two photons. One way to ensure indistinguishability
is to employ narrowband filters to eliminate the spectral correlation
\cite{ZeiHorWei97,FioVosSha02,CheLeeLia06,KalBlaMmo06,FulAliBri07}.
The drawback, however, is that the production rate of the heralded photons becomes low because most usable pairs are filtered
out. This problem can not be overcome by simply increasing the pump power in the photon-pair source, since doing so will lead to strong background noise arising from multipair emission. A solution to this dilemma was proposed by Grice \emph{et al.} \cite{spectral-disentanglement01}, who suggested the creation of photon pairs in spectrally-factorable states so that measuring the signal does not disclose any spectral information about the idler. Since the narrowband filters are abandoned, doing so can dramatically improve the single-photon production rate. With such motivation, extensive
efforts have been made to study the creation of spectrally factorable photon
pairs \cite{Taloring-FWM07,Single-Photon-PTP10} and a few experimental demonstrations have been made
\cite{Single-Photon-PDC07,Herald-Single-Photon-08,Single-Photon-Fiber09,Tailoring-FWM09,TailoringFWM10}.
However, progress has been limited due to the practical difficulty of simultaneously achieving appropriate phase matching and group-velocity matching.

To overcome the experimental limitations, we propose an approach that does not rely on spectral factorability of the photon pairs. Based on a multimode theory, we show that non-factorable states can also lead to high-purity heralded single photons without the use of narrowband filters. The only requirement is that the detection of the signal photons must correspond to a single-mode measurement, which can be achieved by using a broadband filter and an on/off detector whose measurement window is shorter than the coherence time of the signal photons. Since narrowband filters are not used, the production rate can be high while at the same time the noise due to multipair emission remains low. Hence, our approach could lead to high-quality single photons over a broad range of spatiotemporal modes using conventional fibers or crystalline media. In contrast, schemes utilizing factorable states are restrictive because of the requirement for tailorable dispersion properties of the nonlinear media
\cite{Taloring-FWM07,Single-Photon-PTP10,Single-Photon-PCF05,Single-Photon-PCF08,Single-Photon-Fiber09,Single-Photon-fiber09-2}.

We present our multimode theory in section \ref{model}. Using this theory, in section \ref{analysis} we compare the performance of conventional heralding of single photons employing spectrally factorable two-photon states with our proposed scheme to use non-factorable states but with detection working in the single-mode measurement regime. We then give a concrete example of heralding single photons with use of a non-factorable state in section \ref{example}, and finally conclude in section \ref{conclusion}.

\section{The Model}
\label{model}
We consider a simple yet general model, wherein pairs of signal and idler photons are generated in either a $\chi^{(2)}$ waveguide or a $\chi^{(3)}$ fiber. The signal photons are passed through a filter and then measured by an on/off detector. As the detector clicks, idlers photon are heralded. For simplicity, we assume a rectangular-shaped filter with a bandwidth
$B$ and a measurement window $T$, which is shorter than or equal to the inherent time resolution of the detector. Such $T$ can be achieved, for example, by placing a temporal shutter in front of the detector \cite{HalAltKum10}.


For concreteness, we assume the output of the photon-pair source to be in a noiseless state containing at most one photon-pair per pump pulse. Linearizing around phase-matching frequencies, the output state can be written as \cite{Taloring-FWM07}
\begin{eqnarray}
\label{psisi}
   & & |\Psi\rangle=\mathcal{N}\bigg[|\mathrm{vac}\rangle+ \\
    & & \kappa\int d\omega_s ~d\omega_i~ \Phi(\omega_s,\omega_i)\exp\left(i\frac{\mu_s \omega_s+\mu_i\omega_i}{2}\right) |\omega_s,\omega_i\rangle\bigg]. \nonumber
\end{eqnarray}
Here $\mathcal{N}$ is a normalization factor, $\kappa$ is proportional to the pair-generation efficiency, and $\omega_s$ and $\omega_i$ are the angular-frequency detunings of the signal and idler photons, respectively, from the perfect phase-matching point. $|\mathrm{vac}\rangle$ is the vacuum state and $|\omega_s,\omega_i\rangle\equiv\hat{a}^\dag(\omega_s) \hat{a}^\dag(\omega_i)|\mathrm{vac}\rangle$ is the basis for two-photon states, where $\hat{a}^\dag (\omega_s)$ and $\hat{a}^\dag(\omega_i)$ are the creation operators, respectively, for the signal and idler fields, satisfying $[\hat{a}(\omega),\hat{a}^\dag(\omega')]=2\pi \delta(\omega-\omega')$. $\Phi(\omega_s,\omega_i)$ is the joint spectral wavefunction of the two photons that is determined by the pump spectral profile and the phase-matching property of the nonlinear medium in the source. For a Gaussian-profile pump (or two pumps in case of nondegenerate four-wave scattering), one obtains to leading order
\begin{eqnarray}
\label{phisi}
\Phi(\omega_s,\omega_i)=\exp\left(-\frac{(\omega_s+\omega_i)^2}{2\sigma^2}\right)
  \mathrm{sinc}\left(\frac{\mu_s \omega_s+\mu_i\omega_i}{2}\right),
\end{eqnarray}
where $\sigma$ is the pump bandwidth, and $\mu_{s,i}$ are the phase-matching coefficients determined by the dispersion relation and the length of the nonlinear medium \cite{spectral-disentanglement01,Taloring-FWM07}. In Eq.~(\ref{psisi}), the exponential phase term represents the group-velocity dispersion of the signal and idler photons relative to the pump(s). Also from Eq.~(\ref{psisi}), the probability to generate a photon-pair per pump pulse is
\begin{equation}
 \mathcal{P}_\mathrm{pair}=\left[1+1/(4\pi^2\kappa^2 \int d\omega_s d\omega_i |\Phi(\omega_s,\omega_i)|^2)\right]^{-1}.
\end{equation}


We now model the detection stage. Existing studies have decomposed the two-photon wavefunction onto the Schmidt
modes \cite{spectral-disentanglement01,Taloring-FWM07,Herald-Single-Photon-08,
TailoringFWM10}. We argue, however, that such modes have little
physical meaning as they are usually not the eigenmodes of the
detection apparatus. Instead, an appropriate set of modes should be chosen by following the standard procedure for the detection of band-limited signals over the measurement time \cite{PrSp61,ZhuCav90}. In the frequency domain, one-photon states for such modes are given by \cite{SasSuz06}
\begin{equation}
    |m\rangle_s=\frac{1}{2\pi}\int^{B/2}_{-B/2} d\omega_s \phi_m(c,\omega_s) |\omega_s\rangle,~ m=0,1,\ldots,
\end{equation}
where $c=B T/4$, and the wavefunction
\begin{equation}
  \phi_m(c,\omega_s)=\sqrt{\frac{2\pi(2m+1)}{B}} S_{0m}(c,\frac{2\omega_s}{B})
\end{equation}
with $S_{nm}(x,y)$ being the angular prolate spheroidal function. The corresponding eigenvalue
\begin{equation}
 \chi_m(c)=\frac{2 c}{\pi} \left(R^{(1)}_{0m}(c,1)\right)^2\le 1,
\end{equation}
where $R^{(1)}_{nm}(c,x)$ is the radial prolate spheroidal function. Ordering $\chi_0(c)>\chi_1(c)>\chi_2(c)\cdots$,
$m=0$ represents the fundamental detection mode of our interest.

From the above description, the positive operator-valued measure (POVM) for registering a signal photon is given by $\hat{\mathrm{P}}_\mathrm{{on}}=\sum_m \eta_{m} |m\rangle_{ss}\langle m|$, where $\eta_m=\eta \chi_m(c)$ with $\eta$ as the inherent quantum efficiency of detection.
For simplicity, we assume ideal detection with $\eta=1$; generalization to non-ideal detection is straightforward and will be presented elsewhere. The probability for detecting a signal photon is
\begin{eqnarray}
    \mathcal{P}_s &=&\mathrm{Tr}\{\hat{\mathrm{P}}_\mathrm{on} |\Psi\rangle \langle \Psi|\}_{s,i}\\
    &=&\mathcal{P}_\mathrm{pair} \frac{\sum_{m} \eta_{m} \int d\omega_i|\Phi_{m}(\omega_i)|^2 }{2\pi \int d\omega_s d\omega_i |\Phi(\omega_s,\omega_i)|^2}, \nonumber
\end{eqnarray}
where the trace is carried over both the signal and idler states. $\Phi_{m}(\omega_i)$ is the collapsed wavefunction for the idler, given as
\begin{equation}
\label{phim}
    \Phi_{m}(\omega_i)=\int^{B/2}_{-B/2} d\omega_s \phi_{m}(c,\omega_s) \Phi(\omega_s,\omega_i).
\end{equation}
The detection efficiency of the signal, i.e., the probability to detect an emitted signal photon, is then given by $\mathcal{D}_s=\mathcal{P}_s/\mathcal{P}_\mathrm{pair}$. Upon clicking of the detector, an idler photon
is heralded, whose density matrix is given by $\hat{\rho}_i=\mathrm{Tr}\{\hat{\mathrm{P}}_\mathrm{on} |\Psi\rangle \langle \Psi|\}_{s}/\mathcal{P}_s$, where the trace is carried over the signal states. After some algebra, we obtain
\begin{equation}
\label{rhoi}
    \hat{\rho}_i=\frac{\sum_{m} \eta_{m} \int d\omega_i ~ d\omega_i'~ \Phi_m(\omega_i) \Phi^\ast_m(\omega_i') ~|\omega_i\rangle\langle\omega_i'|}{(2\pi)^2\sum_{m} \eta_{m} \int d\omega_i|\Phi_{m}(\omega_i)|^2}.
\end{equation}

\section{Quality of Heralded Photons}
\label{analysis}
In the above section, we have developed a multimode theory to model the heralding of single photons using photon-pair sources based on $\chi^{(2)}$ waveguides or $\chi^{(3)}$ fibers. In this section, we use this theory to analyze the quality of the heralded photons under various spectral-factorability conditions.

To characterize properties of the heralded photons, we expand $\hat{\rho}_i$ onto its eigenstates, $\hat{\rho}_i=\sum^\infty_{n=0} \lambda_n |n\rangle_{ii}\langle n|$, with ordered eigenvalues $\{\lambda_i\}$ such that $\lambda_0>\lambda_1>\lambda_2\cdots $. One characteristic is the heralding efficiency $\mathcal{H}$, which we define as the probability to find the idler photon in state $|0\rangle_i$ so that $\mathcal{H}=\lambda_0$. We note that there is an alternate definition of heralding efficiency, which is the probability that an idler photon can be detected upon clicking of the signal detector. However, that definition does not capture the quality of the heralded photons and some measurement giving the purity of the heralded state must be specified. In contrast, the heralding efficiency $\mathcal{H}$ defined above captures both merits.

Another characteristic is the single-photon production rate. With detection efficiency $\mathcal{D}_s$ and a pair-generation probability $\mathcal{P}_\mathrm{pair}$, the highest achievable production rate is given by $(\mathcal{D}_s/T_\mathrm{min})\mathcal{P}_\mathrm{pair}$. Here, $T_\mathrm{min}$ is the minimum time required for one heralding cycle, which is given by the largest of the following: the
detection window $T$, the temporal length of the pump pulses (taken as $4/\sigma$ in this paper \cite{pulse-def}), the temporal length of the signal photons after the filter, and the underlying pulse length of the state $|0\rangle_i$. As $\mathcal{P}_\mathrm{pair}$ is directly related to the background-noise level due to multipair emission, we define an ``absolute'' production rate  $\mathcal{R}_\mathrm{abs}=\mathcal{D}_s/T_\mathrm{min}$ for the purpose of comparing different heralding scenarios at the same multpair-noise level.

To achieve high heralding efficiency, i.e., $\mathcal{H}\approx 1$, one way is to use photon pairs that are spectrally uncorrelated over the filter band. Indeed, if $\Phi(\omega_s,\omega_i)$ can be written as a product of two independent spectral functions of the signal and idler photons over $[-B/2,B/2]$, the collapsed wavefunctions $\{\Phi_m(\omega_i)\}$ for the idler photons will have the same shape. As a result, the density matrix $\hat{\rho}_i$ will simply represent a pure state, so that $\mathcal{H}=1$. Based on this principle, earlier experiments for heralded photons of high purity were implemented with narrowband filters \cite{KalBlaMmo06,CheLeeLia06,FulAliBri07,Herald-Single-Photon-08}.
They are, however, subjected to fundamentally low production rates. To overcome this difficulty, recent schemes have relied on creating spectrally factorable two-photon states to get rid of the narrowband filters \cite{spectral-disentanglement01, Taloring-FWM07,Single-Photon-PDC07,Single-Photon-PTP10,Single-Photon-PCF05,Single-Photon-PCF08,Single-Photon-Fiber09,Single-Photon-fiber09-2}.

\begin{figure}
\centering \epsfig{figure=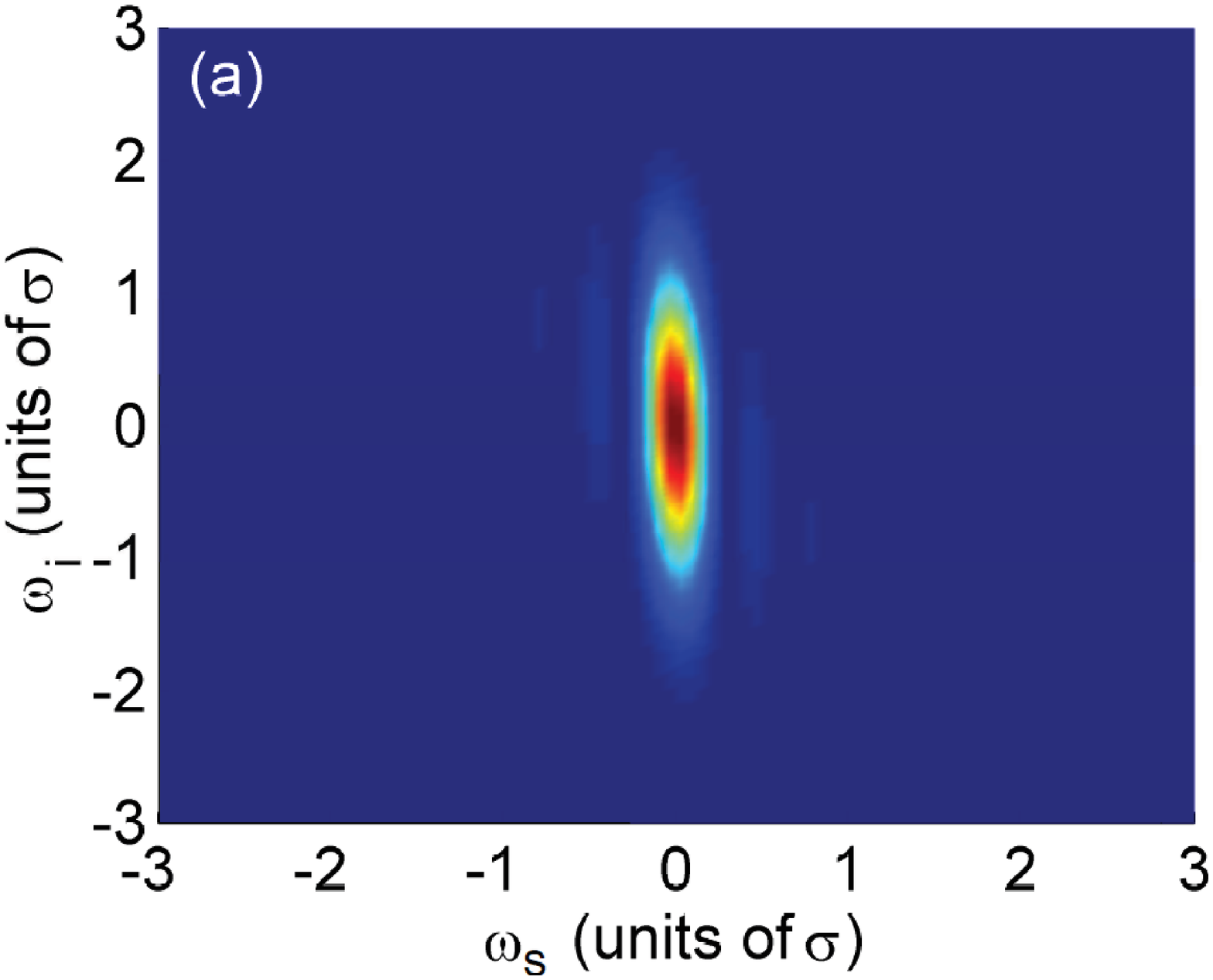, width=6.5cm} \epsfig{figure=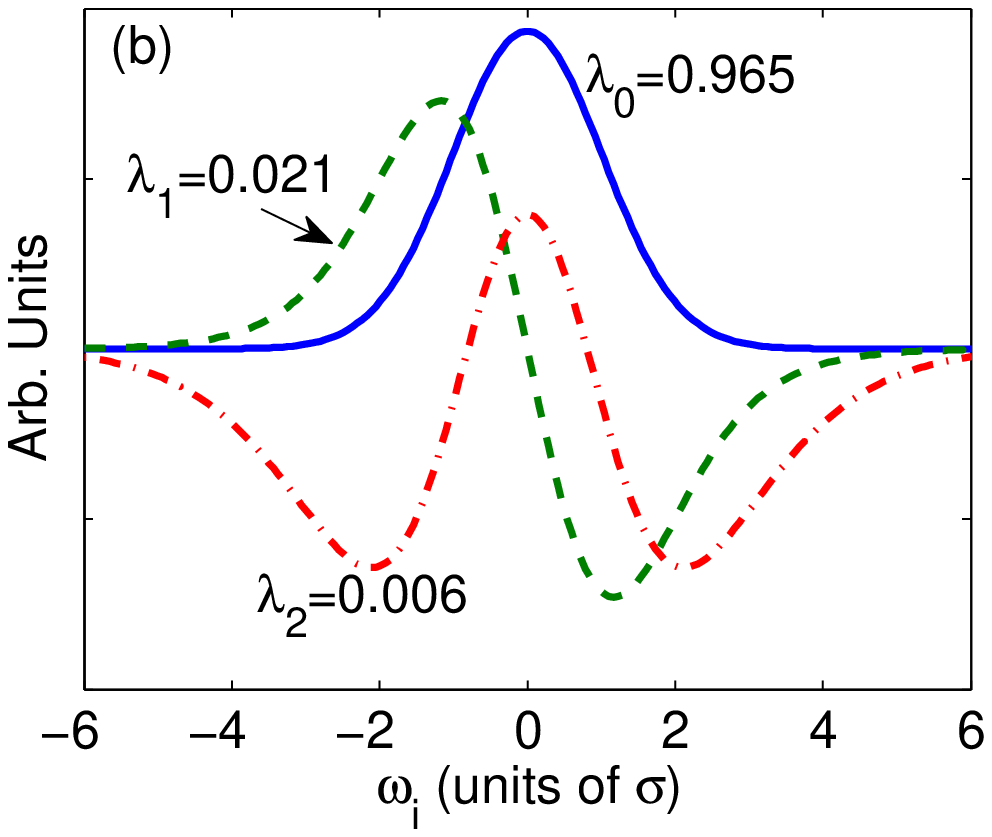, width=7.0cm}\caption{(Color online) (a) Joint two-photon spectrum $|\Phi(\omega_s,\omega_i)|^2$. (b) First three eignmodes of
the heralded idler photons. Parameters: $\mu_s=20/\sigma$, $\mu_i=0$, $B=4\pi\sigma$, and  $T=40/\sigma$.
\label{fig1}}
\end{figure}
As an example, we analyze a widely-adopted scheme using factorable states, where the two-photon spectrum is tailored to correspond to an ellipse
\cite{Herald-Single-Photon-08,Tailoring-FWM09,Single-Photon-Fiber09,TailoringFWM10}.
This can be achieved with $|\mu_i|\ll 1/\sigma$ and $|\mu_s| \gg 1/\sigma$, for which
$
  \Phi(\omega_s,\omega_i)\approx \exp\left(-\frac{\omega_i^2}{2\sigma^2}\right)
  \mathrm{sinc}\left(\frac{\mu_s \omega_s}{2}\right).
$
As a result, we have approximately $\Phi_{m}(\omega_i)\propto \phi_{m}(c,0) \exp\left(-\frac{\omega_i^2}{2\sigma^2}\right)$, giving $\mathcal{H}\approx 1$. Intuitively, this can be
understood by noting that traveling at the same group velocity as the pump, the idler-photon amplitude is forced to be in a single temporal mode---the pump mode. We show the performance of such a heralding source in Fig.~\ref{fig1} for $\mu_s=20/\sigma$, $\mu_i=0$, $B=4\pi\sigma$, and  $T=40/\sigma$. In Fig.~\ref{fig1}(a) we plot the joint two-photon spectrum $|\Phi(\omega_s,\omega_i)|^2$, where a vertical ellipse along the $\omega_i$ axis is shown. The wavefunctions of the first three eigenstates of the heralded photons are shown in Fig.~\ref{fig1}(b). The heralding efficiency is $\mathcal{H}=0.965$, and the detection efficiency of the signal photons is $\mathcal{D}_s=0.997$. Here, the deviation of $\mathcal{H}$ from unity reflects the fact that the two-photon state is not perfectly factorable. For a higher heralding efficiency, one can use a larger $\mu_s$ to improve the factorability. For example, for $\mu_s=40/\sigma$ and $T=80/\sigma$ while keeping the remaining parameters the same, we obtain $\mathcal{H}=0.983$ and $\mathcal{D}_s=0.998$.

The drawback of the above scheme is that the signal photons have a long coherence time ($\sim |\mu_s|$) compared to the pump ($\sim 1/\sigma$). This restricts the duty cycle of the pump(s). Also, the measurement window $T$ must be long ($\sim |\mu_s|$) so as to efficiently detect the signal photons. In effect, one has to wait for a relatively long time before a second short-duration photon can be heralded. This fundamentally limits the repetition rate of the heralding process, leading to a low production rate. In the example shown in Fig.~\ref{fig1}, the absolute production rate is only $\mathcal{R}_\mathrm{abs}=0.025\sigma$. Of course, the repetition rate can be increased by shortening $T$. However, the detection efficiency of the signal photons will be lowered. As a result, the production rate remains more or less the same. We note that this problem can not be overcome by reversing the roles of the signal (long-duration) and idler (short-duration) photons.

\begin{figure}
\centering \epsfig{figure=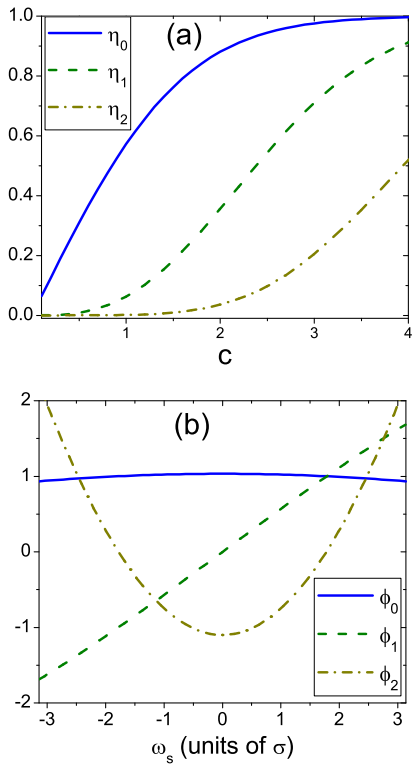, width=5.5cm} \caption{(Color online)  (a) $\eta_0$, $\eta_1$, and $\eta_2$ as functions of $c$. (b) Profiles of the first three detection modes for $B=2\pi$ and $T=0.5$ (corresponding to $c=\pi/4$).  \label{fig2}}
\end{figure}
In the above, we have shown that the scheme using factorable states results in a fundamentally lower production rate of heralded photons. Now we present a different approach that yields a comparable or even higher heralding efficiency at a similar production rate, and which can even use non-factorable states. This approach requires that the filtering and detection lead to a single-mode measurement. This is achieved when $\eta_0\gg\eta_1, \eta_2, \cdots$ for any two-photon state, as shown in (\ref{rhoi}). Here, the key point is that $\{\eta_m\}$ depend only on the product of $B$ and $T$, not individually on them. Hence, even a broadband filter can lead to a single-mode measurement over a sufficiently short detection window. As an example, in Fig.~\ref{fig2}(a), we show $\eta_{0,1,2}$ as a function of $c\equiv BT/4$. For $c<1$, we have $\eta_0\gg \eta_1,\eta_2$, giving rise to approximately a single-mode measurement. We emphasize that this is true for any $B$, as long as $T<4/B$. To understand this, recall that the on/off detector has a time resolution wider than the measurement window $T$. A detection event announces the arrival of a signal photon at an unknown time within the window $T$. In the frequency domain, this corresponds to a detection resolution of $1/T$. Given $c<1$ or $1/T>B/4$, the detector is thus unable to, even in principle, reveal the frequency of the signal photon, resulting in a single-mode measurement. This can be seen in Fig.~\ref{fig2}(b), where the fundamental detection mode has a nearly flat profile over the filter band $[-B/2, B/2]$. Lastly, since $T<4/B$ is required, the detection efficiency of the signal photons will be sub-unity, but not significantly less than one. Note that even for single-mode measurements ($BT<4$), very low $B$ or very low $T$ can lead to pure heralded photons but with low production rates. However, for a properly chosen $B$ and $T$, the repetition rate can still be high, leading to a relatively large production rate for heralded photons of high purity.

\begin{figure}
\centering \epsfig{figure=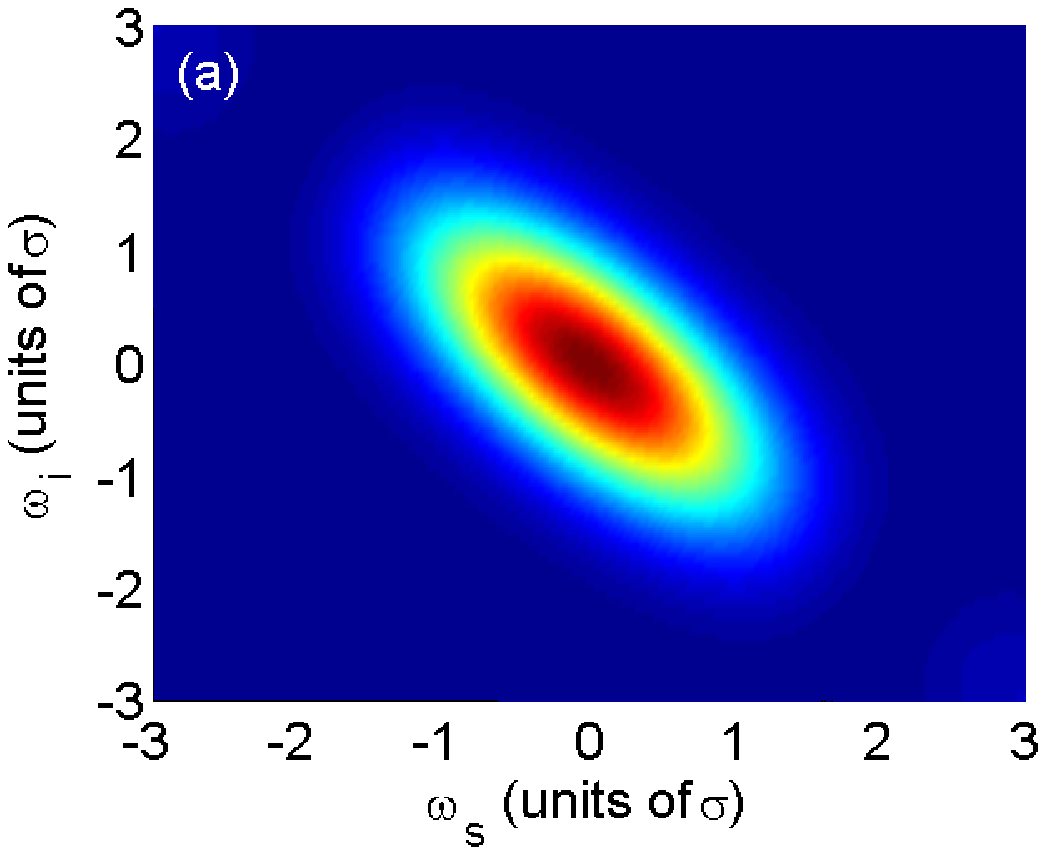, width=7.0cm}
 \epsfig{figure=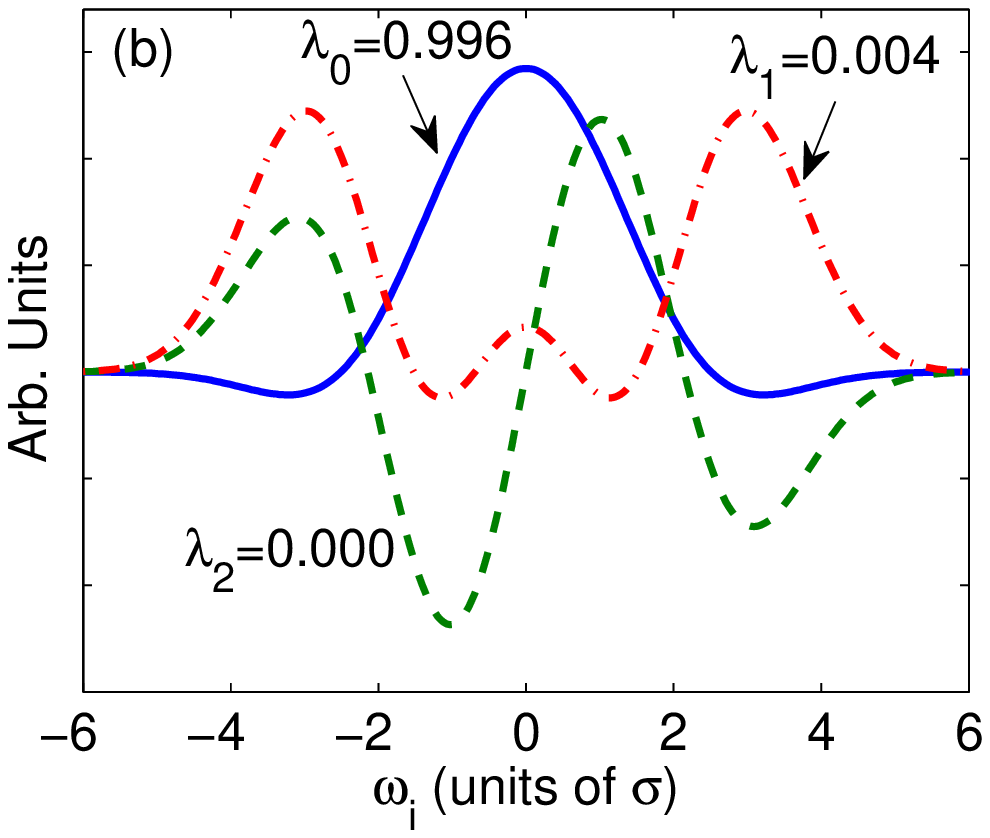, width=7.0cm} \caption{(Color online) Similar to Fig.~\ref{fig1} but for a non-factorable two-photon state and a single-mode measurement. Parameters: $\mu_s=2/\sigma$, $\mu_i=-1/\sigma$, $B=2\pi\sigma$, and $T=0.5/\sigma$. \label{fig3}}
\end{figure}
To make the above analysis concrete, we present an example with the following parameters: $\mu_s=2/\sigma$, $\mu_i=-1/\sigma$, $B=2\pi\sigma$,
and $T=0.5/\sigma$. The underlying two-photon state is clearly non-factorable, as shown in Fig.~\ref{fig3}(a). Yet with
$c=0.79<1$, the detection is approximately single-mode, giving rise to heralded photons of high purity, as shown Fig.~\ref{fig3}(b). The heralding efficiency turns out to be $\mathcal{H}=0.996$, almost one. Of course, this comes at the price of a somewhat lower detection efficiency as discussed above. In this example, $\mathcal{D}_s=0.206$ compared to $0.997$ for the case of a factorable state shown in Fig.~\ref{fig1}. In terms of the production rate, however, this relatively smaller $\mathcal{D}_s$ can be well compensated by a much higher repetition rate. As a result, the absolute production rate can be larger than that using factorable states. For the above parameters, we have $\mathcal{R}_\mathrm{abs}=0.056\sigma$ compared to $0.025\sigma$ for the example shown in Fig.~\ref{fig1}.

\begin{figure}
\centering \epsfig{figure=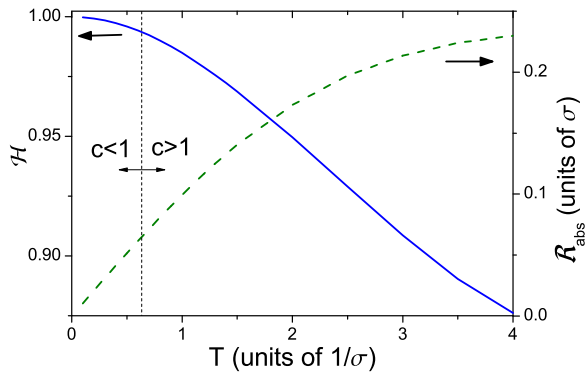, width=8.0cm} \caption{(Color online)  The
heralding efficiency $\mathcal{H}$ and the absolute production rate
$\mathcal{R}_\mathrm{abs}$ as functions of the detection window $T$. All remaining parameters are the same as in Fig.~\ref{fig3}. \label{fig4}}
\end{figure}
For non-factorable states, the heralding efficiency $\mathcal{H}$
decreases with $T$. The absolute production rate $\mathcal{R}_\mathrm{abs}$ is given by the ratio of the signal-photon detection efficiency $\mathcal{D}_s$ and the operation time per heralding cycle $T_\mathrm{min}$. For heralding using broadband filters and nonlinear media of normal dispersion properties, $T_\mathrm{min}$ is usually determined by the pump pulse length, in which case $T<T_\mathrm{min}$ and thus $\mathcal{R}_\mathrm{abs}\propto \mathcal{D}_s$. As $\mathcal{D}_s$ increases with $T$, $\mathcal{R}_\mathrm{abs}$ will also increase with $T$. Hence, there is a fundamental trade-off
between $\mathcal{H}$ and $\mathcal{R}_\mathrm{abs}$ when $T<T_\mathrm{min}$. To illustrate this, we plot $\mathcal{H}$ and $\mathcal{R}_\mathrm{abs}$ as functions of $T$ in Fig.~\ref{fig4}. As shown, the heralding efficiency $\mathcal{H}$ remains above $0.99$ until $T$ increases to a point (corresponding to $c>1$) where the detection is no longer a single-mode measurement. This is because for non-factorable two-photon states, multimode detection will project the idler photons onto a mixture of states, thus lowering the heralding efficiency. On the other hand, the absolute production rate $\mathcal{R}_\mathrm{abs}$ increases as $\mathcal{H}$ decreases. For $c=1$, $\mathcal{R}_\mathrm{abs}=0.065\sigma$, which is over twice larger than is typical with factorable states. Such trade-off
hints that in practice one could tune $T$ to find the optimal
performance for a specific quantum application.

\section{A concrete example}
\label{example}
In this section we study a realistic heralding scheme with use of a 500-m long, liquid-nitrogen cooled, standard single-mode fiber. Following our previous entanglement experiment using such a fiber \cite{HalAltKum09}, we employ Gaussian-shaped pump pulses centered at 1305.0 nm with a transform-limited bandwidth of 0.03 nm. For the signal-photon filter, we choose the central transmission wavelength to be 1306.5 nm and the transmission bandwidth to be 0.14 nm (corresponding to $B/2\pi=25$ GHz). To obtain a short detection window as required by the single-mode-measurement condition, we utilize an all-optical fiber-based switch as a time gate in front of the on/off detector \cite{HalAltKum10}. Such a device has allowed us to switch single photons at ultrahigh speed with low loss, and without disturbing their quantum state \cite{HalAltKum102}. It has the feature that the switching window, thus the detection window $T$ in our heralding setup, can be tuned by adjusting the fiber length of the Sagnac loop used in the switch. For instance, for the parameters used in \cite{HalAltKum102}, $T=180$ ps and $9$ ps can be achieved by using loop lengths of $100$ m and $5$ m, respectively.

\begin{figure}
\centering \epsfig{figure=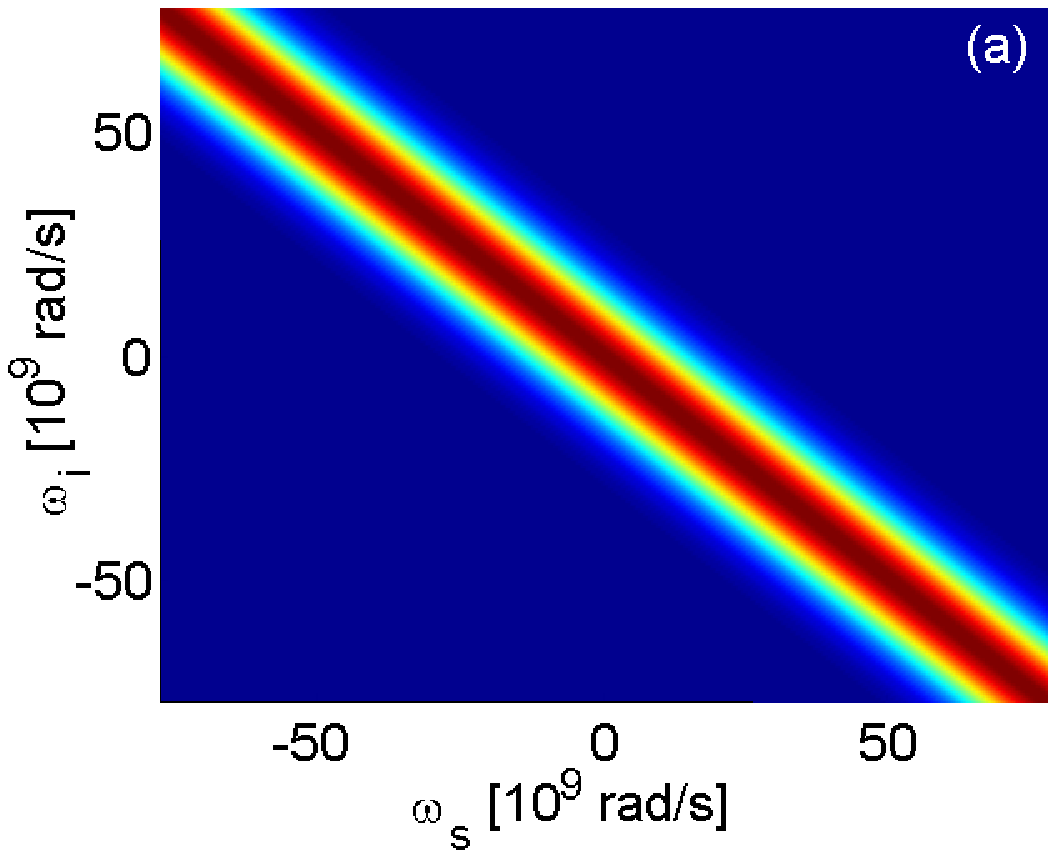, width=7.0cm}
\epsfig{figure=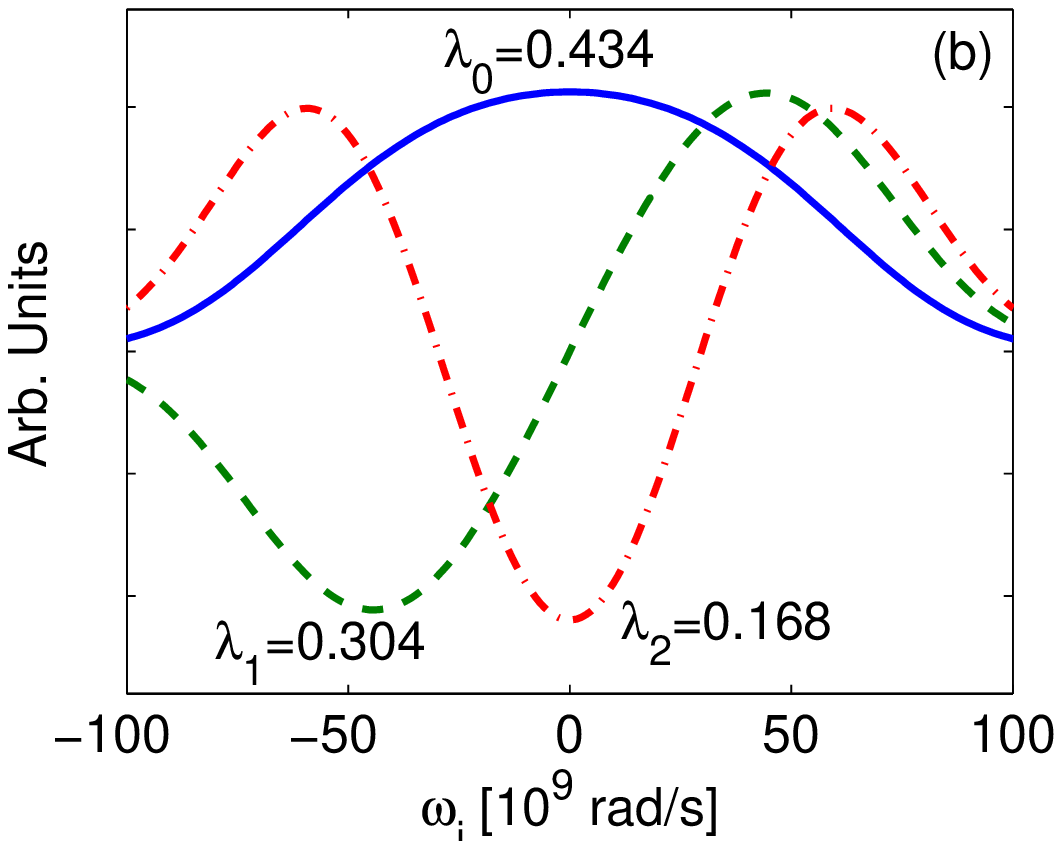, width=7.0cm}
\epsfig{figure=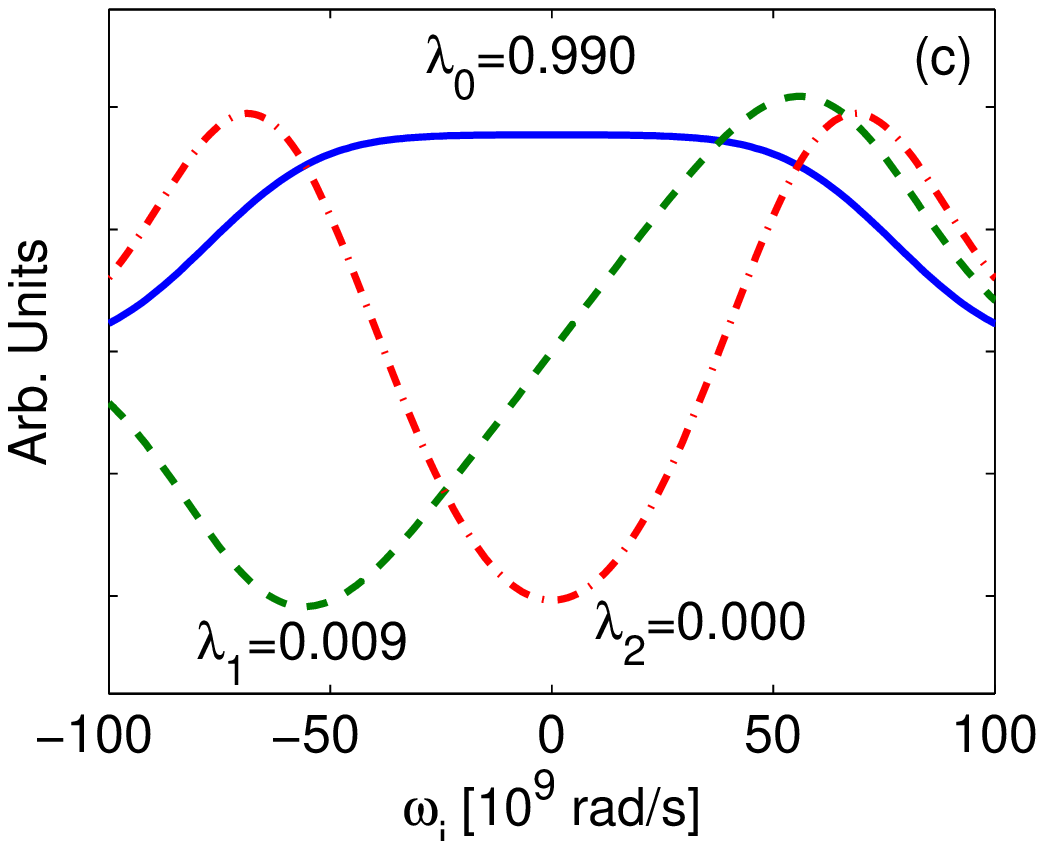, width=7.0cm}
\caption{(Color online) (a) Joint two-photon spectrum $|\Phi(\omega_s,\omega_i)|^2$ for the example in section \ref{example}. (b) and (c) First three eigenmodes of the heralded photons for $T$=180 ps and 9 ps, respectively. The nonlinear medium is a 500-m-long piece of standard single-mode fiber pumped at 1305.0 nm. \label{fig5}}
\end{figure}

In Fig.~\ref{fig5}, we plot the joint two-photon spectrum (a) and the first three eigenmodes of the heralded single photons for the two values of $T$, (b) and (c), respectively. Figure \ref{fig5}(a) clearly shows that the two-photon joint spectrum of the quantum state is non-factorable. For such a state, when $T=180$ ps, the idler photons are heralded in mixed states, as shown by the eigenvalues in Fig.~\ref{fig5}(b). The heralding efficiency $\mathcal{H}$ is only $0.434$. This is because for $B/2\pi=25$ GHz and $T=180$ ps, $c=7>1$, so that the detection apparatus corresponds to a multi-mode measurement, giving rise to a highly impure (i.e., mixed) quantum state for the heralded photons. To achieve a high heralding efficiency and a correspondingly pure state for the heralded photons, we need to shorten the measurement window to realize a single-mode measurement. This is accomplished by using $T=9$ ps, for which $c=0.35<1$. For such a $T$, the eigenstates of the heralded idler photons and the corresponding eigenvalues (occupation probabilities) are shown in Fig.~\ref{fig5}(c). We see that the heralded photons are in a single mode with near-unity probability and the heralding efficiency $\mathcal{H}=0.990$.

For the parameters used in Figs.~\ref{fig5} (b) and (c), the absolute pair production rates turn out to be $\mathcal{R}_\mathrm{abs}=4.644$ GHz and $0.430$ GHz, respectively. To connect these numbers with achievable production rates of the heralded photons in practice, we assume a total detection efficiency (including propagation losses) of $5\%$ for the signal photons. For strong pump pulses with a peak power of 600 mW, which gives a per-pulse pair-production probability of $\mathcal{P}_\mathrm{pair}=0.14$ \cite{HalAltKum09}, the production rates of the heralded photons turn out to be $32.5$ MHz and $3.0$ MHz for $T=180$ ps and 9 ps, respectively. For weak pump pulses with a $160$-mW peak power, for which $\mathcal{P}_\mathrm{pair}=0.015$, the resulting production rates are $3.5$ MHz and $0.3$ MHz, respectively, for the two choices of $T$.

Finally, we note that single photons of even higher quality can be produced by using the same fiber source of photon pairs, but by slightly modifying the experimental setup. For example, by using 0.12-nm-bandwidth transform-limited pump pulses of 160 mW peak power, while keeping all other parameters the same, we can achieve a production rate of $3.5$ MHz with $\mathcal{H}=0.998$ (In practice, $\mathcal{H}$ will be reduced somewhat due to the background noise of spontaneous Raman scattering in the fiber). Hence, our proposed scheme can allow MHz-rate generation of high-quality single photons from a conventional single-mode fiber.
In comparison, the state-of-the-art production rate of heralded photons using spectrally-factorable two-photon states, which are created in a photonic-crystal fiber, is estimated to be $\sim 0.1$ MHz for a heralding efficiency of $\mathcal{H}\sim 0.8$ \cite{TailoringFWM10}.

\section{Conclusion}
\label{conclusion}
We have developed a multimode
theory to model sources of single photons based on heralding. We find that neither spectral factorability nor narrowband filtering is required to achieve both a high heralding efficiency (hence high purity) and a high production rate. Instead, any non-factorable state can be used to herald pure photons as long as the detection approximates a single-mode measurement. The production rate can be higher than the existing schemes using factorable two-photon states, since narrowband filtering is not used. We do not wish to imply that non-factorable states outperform factorable states in all aspects. Rather, our point is that experimental complications, such as spectrum tailoring, can be circumvented by using only a single-mode measurement. The advantages of our scheme include: a) the implementation is potentially easier; b) it does not require special nonlinear media; c) it can be applied to a broad range of spatiotemporal modes.

\begin{acknowledgments}
This research was supported in part by the DARPA ZOE program (Grant No. W31P4Q-09-1-0014) and the U.S. Army Research Office under the Quantum Imaging MURI (Grant No. W911NF-05-1-0197).
\end{acknowledgments}


\begin{thebibliography}{26}
\bibitem{pulse-def} Assuming that the successive Gaussian-shaped pump pulses can be placed as close as twice their $e^{-1}$ full width.
\end{thebibliography}

\end{document}